\documentclass[aps,reprint,superscriptaddress,nofootinbib,longbibliography]{revtex4-2}
\usepackage{graphicx, color, soul}
\usepackage{xcolor, bbold}      
\usepackage{siunitx}
\usepackage{amsmath, amssymb}
\usepackage{braket}
\usepackage{color,soul,url}
\usepackage{float}
\usepackage{wrapfig}
\usepackage[hidelinks]{hyperref}
\usepackage{microtype} 
\usepackage[]{newtxtext} 
\usepackage[]{newtxmath}

\usepackage{hyperref}
\hypersetup{colorlinks,
	linkcolor={blue!75!black!80!yellow},
	citecolor={blue!75!black!80!yellow},
	urlcolor={blue!75!black!80!yellow}
}

\usepackage{booktabs}
\usepackage{colortbl}

\usepackage{makecell}

\definecolor{SMblue}{rgb}{0.5,0.8,1}

\begin{document}
\title{Direct Observation of Landau Levels in Silicon Photonic Crystals}

\author{Maria Barsukova*}
\affiliation{Department of Physics, The Pennsylvania State University, University Park, PA, USA}
\thanks{These authors contributed equally}

\author{Fabien Gris\'e*}
\affiliation{Department Astronomy \& Astrophysics, The Pennsylvania State University, University Park, PA, USA}
\thanks{These authors contributed equally}

\author{Zeyu Zhang*}
\affiliation{Department of Physics, The Pennsylvania State University, University Park, PA, USA}
\thanks{These authors contributed equally}

\author{Sachin Vaidya}
\affiliation{Department of Physics, The Pennsylvania State University, University Park, PA, USA}
\affiliation{Department of Physics, Massachusetts Institute of Technology, Cambridge, MA, USA}

\author{Jonathan Guglielmon}
\affiliation{Department of Physics, The Pennsylvania State University, University Park, PA, USA}

\author{Michael I. Weinstein}
\affiliation{Department of Applied Physics and Applied Mathematics and Department of Mathematics, Columbia University, New York, NY, USA}

\author{Li He}
\affiliation{Department of Physics and Astronomy, University of Pennsylvania, Philadelphia, PA, USA}

\author{Bo Zhen}
\affiliation{Department of Physics and Astronomy, University of Pennsylvania, Philadelphia, PA, USA}

\author{Randall McEntaffer}
\affiliation{Department Astronomy \& Astrophysics, The Pennsylvania State University, University Park, PA, USA}

\author{Mikael C. Rechtsman}
\affiliation{Department of Physics, The Pennsylvania State University, University Park, PA, USA}
\email{mcrworld@psu.edu}

\date{\today}

\begin{abstract}
We experimentally observe photonic Landau levels that arise due to a strain-induced pseudomagnetic field in a silicon photonic crystal slab.  The Landau levels are dispersive (i.e., they are not flat bands) due to the distortion of the unit cell by the strain.  We employ an additional strain which induces a pseudoelectric potential to flatten them.
\end{abstract}
\maketitle

When electrons are confined to a two-dimensional plane and are subjected to an out-of-plane magnetic field, they move in circular cyclotron orbits as a result of the Lorentz force.  In the quantum domain, this cyclotron motion is quantized, and as a consequence, the electrons' energy spectrum splits into discrete, highly degenerate states called Landau levels.  The integer and fractional quantum Hall effects \cite{tsui1982two, klitzing1980new} arise as a direct result; in the fractional case, it is the high degeneracy of Landau levels (i.e., that they are flat bands) that gives rise to effectively strong electron-electron interactions and leads to the fractionalization of charge.  

In free space, photons do not respond to external magnetic fields because they do not carry charge; yet, when propagating in magneto-optical materials, they may respond indirectly as a result of the material's magnetic response. However, this response is weak at optical frequencies.  In 2012, an approach was put forward for emulating magnetic behavior in photonic systems by inhomogeneously straining a photonic lattice. \cite{rechtsman2013strain}.  This implementation was based on an idea proposed for electrons in graphene, where a strain pattern imposed on the lattice would introduce an effective gauge field at the Dirac point, causing electrons to behave as though there were a strong field present, even in the absence of a real magnetic field \cite{guinea2010energy}.  The effect was later demonstrated by directly observing Landau levels in graphene bubbles, where a strain corresponding to an enormous `pseudomagnetic' field of $300T$ was imposed \cite{levy2010strain}.  Since the original photonic experiment, Landau levels were also proposed and observed in exciton-polariton condensates \cite{jamadi2020direct} and in mechanical systems \cite{abbaszadeh2017sonic, brendel2017pseudomagnetic, peri2019axial}.  Moreover, there have been a number of theoretical proposals for how Landau levels may be used in the context of photonics that are intrinsically distinct from the electronic case \cite{schomerus2013parity,lledo2022polariton}.  

Here, we directly observe Landau levels in two-dimensional silicon photonic crystal slabs in the nanophotonic domain.  Moreover, we go beyond purely pseudomagnetic effects and demonstrate that strains corresponding to {\it pseudoelectric} fields act to flatten the Landau levels that inherit dispersion from the form of the pseudomagnetic strain.    There are several key differences and advantages of pseudomagnetism in photonic crystals compared to previous realizations of photonic pseudomagnetism.  First, photonic crystals have been demonstrated to enhance light-matter interaction via cavity modes and flat bands \cite{li2008systematic,jukic2012flat,yang2023photonic}. This enhancement is generated as a result of the lattice. In contrast, for systems composed of individual, isolated guiding or resonant elements (as in Refs. \cite{rechtsman2013strain, jamadi2020direct}), lattice effects are not leveraged because strong enhancement would occur even in a single site.  Second, besides having unit cells that are an order of magnitude smaller, photonic crystals can in practice have much larger system sizes compared to previous realizations (millions compared to hundreds of unit cells), and can be realized with smaller loss in the silicon platform.  Since Landau level degeneracy scales with system size and the linewidth increases with loss, photonic crystals allow for increased degeneracy and significantly improved spectral resolution of the levels. 

Further, since photonic crystals do not have an associated tight-binding theory, the original theoretical framework relating strain to pseudomagnetism is not directly applicable, rendering a new understanding necessary; the appropriate effective Hamiltonians for strain-dependent emergent parameters for two-dimensional photonic crystals were derived in our previous theoretical work \cite{guglielmon2021Landau}, and are extended to the slab geometry here (2D slab embedded in 3D space). 
Our establishment of a new analytical method of understanding and describing aperiodicity in photonic crystals (i.e., using pseudomagnetic fields) will be useful in their optimization for many different functions; this has traditionally been approached by using direct numerical optimization \cite{men2014robust, vuckovic2002optimization, bendsoe2003topology}.

\begin{figure*}[t]
    \centering
    \includegraphics[width=2\columnwidth]{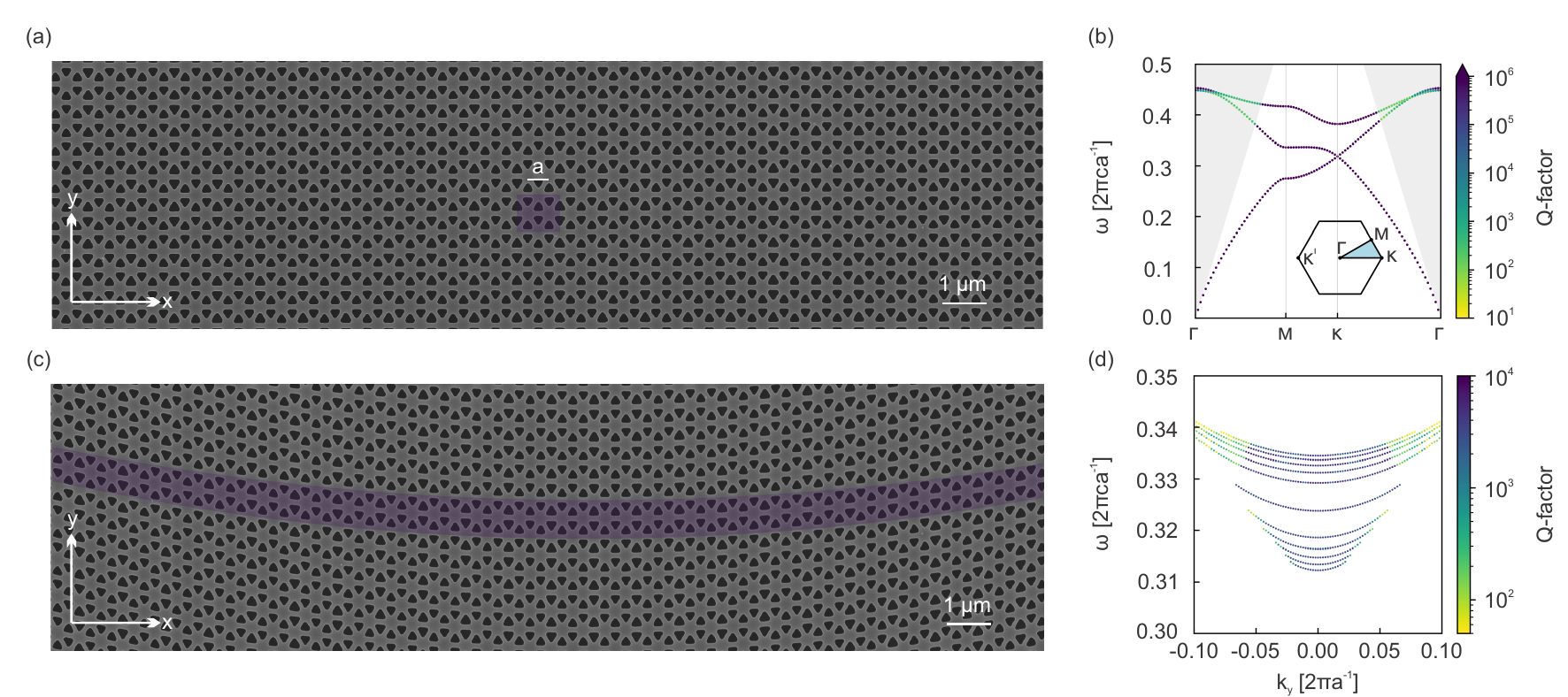}
    \caption{Unstrained and strained photonic crystals and their calculated band structures. (a) Scanning electron microscope image of the unstrained periodic photonic crystal. The period-doubled unit cell is highlighted in purple; $a$, lattice constant without period doubling, is indicated in white. (b) Simulated band structure for the unstrained structure, showing a Dirac point for the TE-like modes.  Color indicates quality factors of the states. (c) SEM image of the strained pattern, in which the strain causes the loss of periodicity in the $x$ direction. The $y$-periodic unit cell of the strained pattern is highlighted in purple. (d) Simulated band structure for the strained pattern ($\kappa=0.0632a^{-1}$), showing the emergence of Landau levels in the vicinity of the Dirac frequency. Only the first eleven Landau levels $(n = -5\text{ to }5)$ are shown.} 
    \label{fig1}
\end{figure*}

Our starting point is a photonic crystal structure consisting of rounded triangular air holes in a silicon slab \cite{barik2016two} that rests on a silica substrate. The holes form an underlying honeycomb pattern with $C_{6v}$-symmetry. As a result, this lattice hosts Dirac points at the $\mathbf{K}$ and $\mathbf{K^{\prime}}$ points in the Brillouin zone \cite{neto2009electronic,fefferman2012honeycomb,Dirac2DPhotonic}. As these Dirac points lie below the light line of vacuum, they are not detectable via free-space excitation. To allow radiative coupling from outside the slab, we introduce a small period-doubling perturbation by changing the size of some of the holes (more details can be found in Supplementary Information Section 2). This makes the unit cell of the lattice rectangular, and the band structure is folded such that the Dirac cone resides along the $k_x$ axis and lies above the light line of vacuum.  A scanning electron microscope image of the structure is shown in Fig.~\ref{fig1}(a); the period-doubled unit cell is shaded in purple.    

We numerically compute the band structure (in the transverse electric polarization) using the guided mode expansion method as implemented in the open-source software package \textsc{Legume} \cite{Legume}. Fig.~\ref{fig1}(b) shows the linearly-dispersing transverse electric (TE)-like bands that exhibit a Dirac point at $\mathbf{K}$, with a frequency $\omega_{\rm D}=0.318\ [2\pi ca^{-1}]$. Here $a$ is the lattice constant of the underlying hexagonal lattice structure and $c$ is the speed of light.  The period-doubling procedure very slightly changes the Dirac frequency (see Supplementary Information Section 2). 

Next, we introduce a strain pattern in our structure by deforming the lattice as shown in Fig.~\ref{fig1}(c). Here, the term \textit{strain } refers not to a strain induced by a physically applied stress, but to the deformation of the dielectric pattern that is directly etched into the silicon. The specific strain pattern is achieved by mapping every point $(x, y,z)$ to $(x, y + a (\kappa x)^2,z)$, where $\kappa$ is the strength of the strain. This deformation breaks periodicity in the $x$ direction, but retains periodicity along the $y$ direction. The spatial scale separation ensured by the assumption of small and slowly varying strain, $\kappa a \ll 1$, allows us to develop a multiple scale \cite{bender1999advanced} variant of degenerate perturbation theory to expand the eigenstates and eigenvalues of the strained system. The eigenstates are, to leading order in $\kappa$, a slow spatial modulation of the degenerate Bloch modes associated with the Dirac point of the unstrained ($\kappa=0$) structure.

The resulting effective Hamiltonian, which incorporates the strain, is given by
\begin{equation}
\mathcal{H}_{\rm eff}=E_{\rm D}\sigma_0+v_{\rm D}\left[\left(-i\frac{\partial}{\partial x}\right)\sigma_1+\left(-i\frac{\partial}{\partial y}+\frac{4ab_*\kappa^2}{v_{\rm D}} x\right)\sigma_2\right],
\label{eq_effective_H}
\end{equation}

\noindent where $E_{\rm D}=\left(\omega_{\rm D}/c\right)^2$, $\sigma_0$, $\sigma_1$ and $\sigma_2$ are Pauli matrices, and $v_{\rm D}=0.915a^{-1}$ and $b_*=0.606a^{-2}$ are two parameters calculated from the modes of the unstrained structure at energy $E_{\rm D}$. A detailed derivation can be found in Supplementary Information section 3, where explicit expressions for $b_*$ and $v_{\rm D}$ in terms of the eigenstates of the periodic structure are displayed. We note that the effective Hamiltonian displayed in Eq. (1) is derived directly from the continuum theory of photonic crystals; this is fundamentally different from the previous work \cite{guinea2010energy} based on the tight-binding approximation. Our approach extends the the methods  of Ref. \cite{guglielmon2021Landau}  to the three-dimensional setting of the slab geometry, where vectorial effects play a role. 

Equation~\eqref{eq_effective_H} corresponds to a two-dimensional Dirac Hamiltonian describing massless spin-$1/2$ relativistic particles under a constant (pseudo)magnetic field pointing in the out-of-plane direction, where the magnetic field has a strength of $B_{\rm eff} = 4ab_*\kappa^2/v_{\rm D}$ and is described by a vector potential in the Landau gauge. The discrete energies that are eigenvalues of the Hamiltonian in Eq.~\eqref{eq_effective_H} for an electron are known as Landau levels. The energy eigenvalue of the $n^{th}$ level is proportional to $\sqrt{|n|}$, where $n$ is an integer. Analogously, for our photonic crystal slabs, the frequency eigenvalues of the electromagnetic eigenmodes are, to first order in $\kappa$, proportional to $\sqrt{|n|}$ and can be expressed as $\omega_n = \omega_{\rm D} \pm(c^2v_{\rm D}/\sqrt{2}\omega_{\rm D})\sqrt{B_{\rm eff} \lvert n\rvert}$, where $n$ is an integer.  

\begin{figure*}[t]
    \centering
    \includegraphics[width=2\columnwidth]{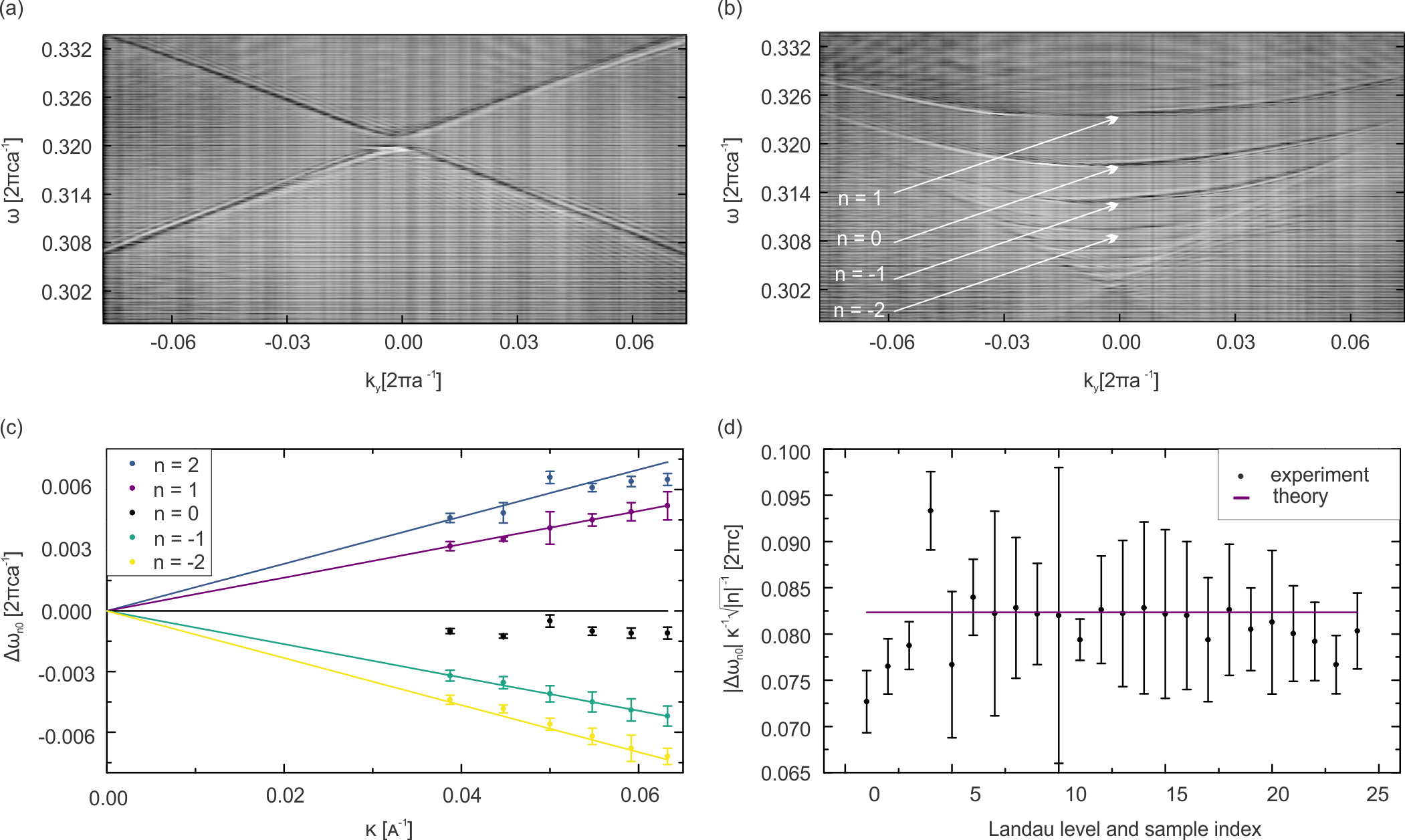}
\caption{Observation of Landau levels in the spectrum of the strained photonic crystal. (a) Experimentally measured band structure of the unstrained honeycomb pattern, showing the Dirac point in the vicinity of \textit{$\omega_{\rm D} = 0.32 ~ [2\pi ca^{-1}]$}.
(b) Measured band structure for the strained pattern. The uniform pseudomagnetic field created by the strain causes the Dirac point to split into sets of discrete Landau levels. Here \textit{$\kappa = 0.0632 a^{-1}$}, as in the numerical results shown in Fig.~\ref{fig1}(d).
(c) Landau level energy spacing ($\Delta \omega_{n,0} = \omega_n - \omega_{0}^{\prime}$) is a linear function of the strain strength $\kappa$. (d) The Landau levels lie at energies proportional to $\sqrt{\lvert n \rvert}$; this behavior is captured by plotting the dimensionless quantity $\frac{\lvert\omega_n - \omega_{0}^{\prime}\rvert} {\sqrt{\lvert n \rvert} \kappa}$ for each Landau level, including six different samples of different strain strengths, $\kappa$.  In both (c) and (d), the solid lines are the theoretical prediction with no free parameters.}
\label{fig2}
\end{figure*}

To corroborate our analytical results given in Eq.~\eqref{eq_effective_H}, we also perform numerical simulations of the strained structure using the guided-mode expansion method. The strain is implemented in a dielectric profile which spans 199 period-doubled unit cells in the $x$-direction. Due to the preservation of lattice periodicity along the $y$-direction, $k_y$ is conserved and the frequencies of the bands can be plotted as functions of $k_y$, as shown in Fig.~\ref{fig1}(d). Here, we observe the splitting of the spectrum near the Dirac point into discrete Landau levels due to the strain-induced pseudomagnetic field, where the spacing of these levels is proportional to $\sqrt{|n|}$ for a fixed value of $\kappa$.

To demonstrate the formation of Landau levels in such a system, we use electron-beam lithography to fabricate both the periodic and the strained patterns in a silicon slab ($\varepsilon = 12.11$) on top of a silica substrate ($\varepsilon = 2.25$). A detailed description of the fabrication methods can be found in Supplementary Material Section 1.  Figures~\ref{fig1}(a) and (c) show scanning electron microscope (SEM) images of the fabricated structures. The structure in Fig.~\ref{fig1}(a) has a periodicity along the $x$ direction of $2a=980~nm$.

To experimentally characterize the photonic bands of these structures, we perform angle- and frequency-resolved reflection measurements. The samples are illuminated by a tunable continuous wave laser (Keysight 81606A) with a wavelength range of $\lambda = 1.45$ - $1.65~\mu m$ ($\pm1.5~pm$ absolute wavelength resolution accuracy), and a laser linewidth coherence control of 10 kHz.  
We measure the iso-frequency contours of the fabricated photonic crystal slabs using back focal plane (BFP) imaging. We then extract the Landau-level band structures by observing the photonic crystal resonances at a fixed $k_x$ corresponding to the location of the Dirac point of the unstrained structure. Details of the experimental setup can be found in Supplementary Material Section 1.

\begin{figure*}[t]
    \centering
   \includegraphics[width=2\columnwidth]{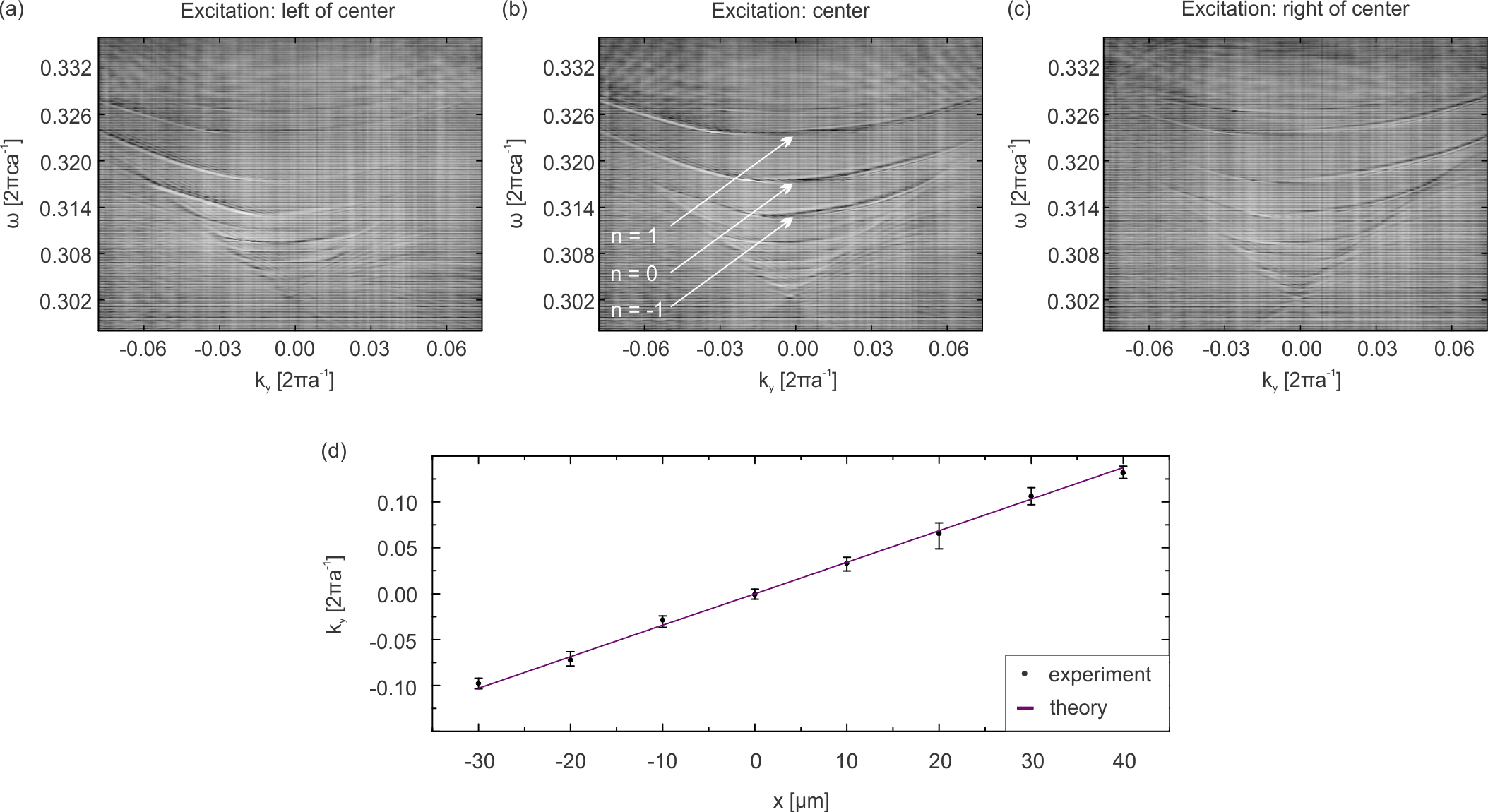}
\caption{Landau levels are excited at increasing $k_y$ in the band structure as the input beam is moved from left to right through the sample.  Experimental band structures showing the excitation of the Landau levels when the beam is injected at (a) $x = -20~\mu$m, (b) $x = 0~\mu$m, and (c) $x = 20~\mu$m.  (d) Center of Landau level excitation in $k_y$ plotted against the position of the input beam, $x$.  The linear relationship is characteristic of Landau level eigenstates in the Landau gauge.  The solid line represents the theoretical prediction, with no free parameters. 
}\label{fig3}
\end{figure*}

Fig.~\ref{fig2}(a) shows the bands of the unstrained structure, obtained by BFP imaging, where we clearly observe linearly dispersing bands near the Dirac point. We note that a small gap is observed at the Dirac point - this is due to inevitable fabrication disorder that breaks inversion symmetry. 
We show in Supplementary Information Section 4 that the breaking of inversion symmetry affects the zeroth Landau level significantly more than the others.

Next, we measure the bands of the strained photonic crystal slabs described above and find the emergence of discrete Landau levels, as shown in Fig.~\ref{fig2}(b).  While the effective theory predicts that the Landau levels should be flat, we see that they are dispersive in both simulation (Fig.~\ref{fig1}(d)) and experiment (Fig.~\ref{fig2}(b)), i.e., the bands are concave-up.  This arises due to the fact that, by adding strain, the unit cell is distorted locally as a function of $x$. This distortion effectively adds a parabolic potential to the Hamiltonian (i.e., $H\sim x^2\sigma_0$), which in turn causes the dispersion of the Landau level bands. A detailed explanation can be found in Supplementary Information Section 5. 

According to the effective theory (Eq. \ref{eq_effective_H}), the $n=0$ level should be at the center of all Landau levels. However, due to the aforementioned inversion-symmetry breaking, this level is slightly shifted away from the center (see Supplementary Information Section 4). As a result, we use a new reference frequency of $\omega_{0}^{\prime} = \frac{1}{2}(\omega_{-1} + \omega_1)$ as the Dirac frequency to calculate the Landau level spacings, defined as $\omega_n - \omega_{0}^{\prime}$. In Fig.~\ref{fig2}(c), we compare the theoretically and experimentally obtained level spacings at $k_y=0 [2\pi a^{-1}]$ under different strain strengths (characterized by $\kappa$) and observe good agreement between the two.  From the experimental data, we also calculate the normalized quantity $|\omega_n - \omega_{0}^{\prime}|/(\kappa \sqrt{|n|})$, which should be a constant for all Landau levels. We again observe good agreement between experiment and the theoretically-predicted value of $0.0823 [2\pi c]$, as shown in Fig.~\ref{fig2}(d).  In both Figs.~\ref{fig2}(c) and (d), the theoretical plots (solid lines) are obtained directly from analytical predictions, and have no free parameters.

\begin{figure*}[t]
    \centering
    \includegraphics[width=2\columnwidth]{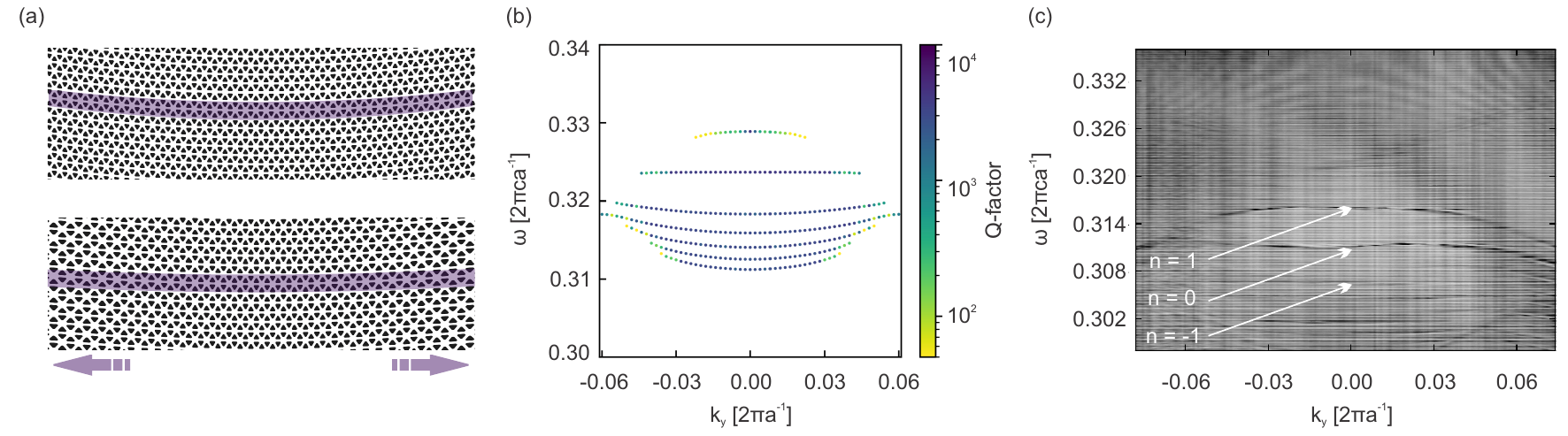}
\caption{The introduction of a pseudoelectric potential acts to flatten the Landau level bands.  (a) Schematic figure depicting the strain profile that corresponds to the pseudoelectric potential used to mitigate the Landau level dispersion. (b) Simulated band structure with $\kappa=0.0632a^{-1}$ and $\beta=0.0364$, where the Landau level $n=0$ is predicted to be flattened.  (c) Experimental data for $\kappa=0.0632a^{-1}$ and $\beta=0.0364$, showing the flattening of the Landau levels.
} \label{fig4}
\end{figure*}

It is clear from Fig. \ref{fig2}(b) that, as $n$ decreases, the range of $k_y$ values over which the $n^{th}$ Landau level is observed becomes smaller. This can be intuitively understood as arising from the interaction of the Landau level states with other states that reside toward the far left and right sides of the sample. These states rise in energy as one moves away from the sample center along the $x$ direction. We know from Eq. (1) that the Landau level states are harmonic oscillator eigenstates centered at $x=k_y/B_{\rm eff}$, with spatial widths of $\Delta x_n=\sqrt{(2|n|+\delta_{0,n})/2B_{\rm eff}}$. As $k_y$ is increased, the Landau level center translates and the tail of the Landau level eventually interacts with the states mentioned above, leading to an increased linewidth. More details are given in Supplementary Information Section 3. 

The fact that the $x$-position of the Landau level state varies linearly with $k_y$ leads to another clear observable: when the input beam is moved from left to right in real space along the $x$-direction, the Landau level states at increasing $k_y$ are selectively excited, and therefore appear more clearly in the band structure. We observe this effect directly, as shown in Fig.~\ref{fig3}(a) through (c): when the input beam is on the left side of the sample (i.e., $x<0$), we see that the modes on the left side of the band structure ($k_y<0$) are more strongly excited, but as the input beam is moved rightward, we observe that the modes on the right side of the band structure are increasingly excited. 

To further study the relationship between $x$ and $k_y$, we extract the boundary in $k_y$-space between the modes that are excited and those that are not excited. For input beams positioned left of center, we extract the right boundary, and for input beams positioned right of center, we extract the left boundary. The boundary values differ from the excitation centers by an overall offset, which we remove by fitting the data to a line and subtracting the intercept (one for the left-boundary data and one for the right-boundary data).  Using this procedure, we obtain the relationship between the Landau level horizontal position and the average vertical momentum, $k_y$, of the excited modes.  The linear relationship between these, as shown in Fig.~\ref{fig3}(d), evidences the direct proportionality between the Landau level positions and $k_y$.  

We next turn our attention to the mitigation of the Landau level dispersion. As explained earlier, Eq.~\ref{eq_effective_H} predicts flat Landau levels. However, in the simulations and experiments, the Landau levels exhibit quadratic dispersion as $k_y$ is varied. As shown in Ref.~\cite{guglielmon2021Landau}, it is possible to mitigate this dispersion by introducing an additional strain profile, which induces a {\it pseudo-electric} potential. Specifically, we add a cubic term to the deformation such that the point $(x, y, z)$ is mapped to $(x + a\beta(\kappa x)^3, y + a(\kappa x)^2,z)$. The parameter $\beta$ controls the strength of this additional strain in the $x$-direction. A schematic of the strained structure, which induces both pseudomagnetic and pseudoelectric fields, is shown in Fig.~\ref{fig4}(a) (further details are given in Supplementary Information Section 5).  The reason why the pseudoelectric field counters the Landau level dispersion, to leading order, can be explained as follows.  To leading order, the form of the pseudoelectric field gives rise to a potential $V_{\rm eff}=3a\beta m \kappa^2 x^2\sigma_0$ (to be added to \eqref{eq_effective_H}) which is similar to that which creates the dispersion in the first place (here $m=-3.28a^{-2}$ is a parameter calculated entirely from the states of the periodic structure).  Since the spatial positions of the Landau level eigenstates grow linearly with $k_y$, a quadratic potential in $x$ is equivalent to a parabolic dispersion in $k_y$. An appropriate choice of the field strength (and sign) will then counteract the original dispersion induced by the strain associated with the pseudomagnetic field.  

By choosing $\beta$ appropriately, the quadratic dispersion of the Landau levels can be mitigated, leading to nearly flat bands. We note that each Landau level requires a different value of $\beta$ to counteract its dispersion. More details are given in Supplementary Information Section 5. Fig.~\ref{fig4}(b) shows numerical simulations of the flattened Landau levels for a structure with pseudomagnetic and pseudoelectric fields induced by a strain with $\kappa=0.0632a^{-1}$ and $\beta=0.0364$. Here, the $n=0$ level is targeted, but other levels are also evidently flatter. Fig.~\ref{fig4} (c) shows the experimental data for a strained structure with the same values of $\kappa$ and $\beta$ given above, where a good agreement is observed between theory and experiment. 

In conclusion, we have directly observed Landau levels in the spectra of two-dimensional silicon photonic crystal slabs.  As in graphene, the Landau level energies are proportional to $\sqrt{|n|}$, where $n$ is an integer.  The Landau level bands are found to be dispersive, which can be explained by a distortion of the unit cell as a result of the strain. We further showed that this dispersion can be mitigated by adding an additional strain that induces a position-dependent pseudoelectric field (i.e., a potential).  Landau levels constitute a new methodology for enhancing light-matter interaction which is distinct from standard slow light or cavity enhancement, because a flat band acts essentially as a `cavity everywhere in space'. The realization of optical pseudomagnetism prompts several new questions and directions of inquiry, including: whether Landau-level flat bands can be used to enhance light-matter coupling more efficiently than conventional photonic crystal flat bands or other points of high degeneracy (such as van Hove singularities); the question of the nature of wave mixing processes such as four-wave mixing among Landau levels; and whether the square-root structure of the eigenvalue spacing can lead to different properties associated with entangled pair or frequency comb generation.  More broadly, the framework of pseudomagnetism gives an analytical handle on aperiodic photonic structures, allowing for a new approach to designing devices and better understanding their behavior.     

\section* {Acknowledgements}
We gratefully acknowledge funding support from the Office of Naval Research MURI program under agreement number N00014-20-1-2325, the Air Force Office of Scientific Research MURI program under agreement number FA9550-22-1-0339, as well as the Kaufman and Packard foundations under grant numbers KA2020-114794 and 2017-66821, respectively.  This research was also supported in part by National Science Foundation grants DMS-1620422 (MCR), DMS-1620418 (MIW), DMS-1908657 (MIW) and DMS-1937254 (MIW), as well as Simons Foundation Math + X Investigator Award \#376319 (MIW). The authors acknowledge the Nanofabrication Lab within the Materials Research Institute at Penn State and the help of Michael Labella, as well as seed funding from the Center for Nanofabricated Optics at Penn State University. F.G. thanks GenISys and, in particular, Roger McCay for his help in optimizing the fracturing of the electron-beam patterns. M.B. thanks Sebabrata Mukherjee and Alexander Cerjan for fruitful discussions in the early stages of the project and help with numerical optimization.

We would like to note that the group of Ewold Verhagen has concurrently posted a similar work on the observation of Landau levels in photonic crystals.

\bibliography{reference}

\begin{thebibliography}{24}%
\makeatletter
\providecommand \@ifxundefined [1]{%
 \@ifx{#1\undefined}
}%
\providecommand \@ifnum [1]{%
 \ifnum #1\expandafter \@firstoftwo
 \else \expandafter \@secondoftwo
 \fi
}%
\providecommand \@ifx [1]{%
 \ifx #1\expandafter \@firstoftwo
 \else \expandafter \@secondoftwo
 \fi
}%
\providecommand \natexlab [1]{#1}%
\providecommand \enquote  [1]{``#1''}%
\providecommand \bibnamefont  [1]{#1}%
\providecommand \bibfnamefont [1]{#1}%
\providecommand \citenamefont [1]{#1}%
\providecommand \href@noop [0]{\@secondoftwo}%
\providecommand \href [0]{\begingroup \@sanitize@url \@href}%
\providecommand \@href[1]{\@@startlink{#1}\@@href}%
\providecommand \@@href[1]{\endgroup#1\@@endlink}%
\providecommand \@sanitize@url [0]{\catcode `\\12\catcode `\$12\catcode
  `\&12\catcode `\#12\catcode `\^12\catcode `\_12\catcode `\%12\relax}%
\providecommand \@@startlink[1]{}%
\providecommand \@@endlink[0]{}%
\providecommand \url  [0]{\begingroup\@sanitize@url \@url }%
\providecommand \@url [1]{\endgroup\@href {#1}{\urlprefix }}%
\providecommand \urlprefix  [0]{URL }%
\providecommand \Eprint [0]{\href }%
\providecommand \doibase [0]{https://doi.org/}%
\providecommand \selectlanguage [0]{\@gobble}%
\providecommand \bibinfo  [0]{\@secondoftwo}%
\providecommand \bibfield  [0]{\@secondoftwo}%
\providecommand \translation [1]{[#1]}%
\providecommand \BibitemOpen [0]{}%
\providecommand \bibitemStop [0]{}%
\providecommand \bibitemNoStop [0]{.\EOS\space}%
\providecommand \EOS [0]{\spacefactor3000\relax}%
\providecommand \BibitemShut  [1]{\csname bibitem#1\endcsname}%
\let\auto@bib@innerbib\@empty
\bibitem [{\citenamefont {Tsui}\ \emph {et~al.}(1982)\citenamefont {Tsui},
  \citenamefont {Stormer},\ and\ \citenamefont {Gossard}}]{tsui1982two}%
  \BibitemOpen
  \bibfield  {author} {\bibinfo {author} {\bibfnamefont {D.~C.}\ \bibnamefont
  {Tsui}}, \bibinfo {author} {\bibfnamefont {H.~L.}\ \bibnamefont {Stormer}},\
  and\ \bibinfo {author} {\bibfnamefont {A.~C.}\ \bibnamefont {Gossard}},\
  }\href@noop {} {\bibfield  {journal} {\bibinfo  {journal} {Physical Review
  Letters}\ }\textbf {\bibinfo {volume} {48}},\ \bibinfo {pages} {1559}
  (\bibinfo {year} {1982})}\BibitemShut {NoStop}%
\bibitem [{\citenamefont {Klitzing}\ \emph {et~al.}(1980)\citenamefont
  {Klitzing}, \citenamefont {Dorda},\ and\ \citenamefont
  {Pepper}}]{klitzing1980new}%
  \BibitemOpen
  \bibfield  {author} {\bibinfo {author} {\bibfnamefont {K.~v.}\ \bibnamefont
  {Klitzing}}, \bibinfo {author} {\bibfnamefont {G.}~\bibnamefont {Dorda}},\
  and\ \bibinfo {author} {\bibfnamefont {M.}~\bibnamefont {Pepper}},\
  }\href@noop {} {\bibfield  {journal} {\bibinfo  {journal} {Physical review
  letters}\ }\textbf {\bibinfo {volume} {45}},\ \bibinfo {pages} {494}
  (\bibinfo {year} {1980})}\BibitemShut {NoStop}%
\bibitem [{\citenamefont {Rechtsman}\ \emph {et~al.}(2013)\citenamefont
  {Rechtsman}, \citenamefont {Zeuner}, \citenamefont {T{\"u}nnermann},
  \citenamefont {Nolte}, \citenamefont {Segev},\ and\ \citenamefont
  {Szameit}}]{rechtsman2013strain}%
  \BibitemOpen
  \bibfield  {author} {\bibinfo {author} {\bibfnamefont {M.~C.}\ \bibnamefont
  {Rechtsman}}, \bibinfo {author} {\bibfnamefont {J.~M.}\ \bibnamefont
  {Zeuner}}, \bibinfo {author} {\bibfnamefont {A.}~\bibnamefont
  {T{\"u}nnermann}}, \bibinfo {author} {\bibfnamefont {S.}~\bibnamefont
  {Nolte}}, \bibinfo {author} {\bibfnamefont {M.}~\bibnamefont {Segev}},\ and\
  \bibinfo {author} {\bibfnamefont {A.}~\bibnamefont {Szameit}},\ }\href@noop
  {} {\bibfield  {journal} {\bibinfo  {journal} {Nature Photonics}\ }\textbf
  {\bibinfo {volume} {7}},\ \bibinfo {pages} {153} (\bibinfo {year}
  {2013})}\BibitemShut {NoStop}%
\bibitem [{\citenamefont {Guinea}\ \emph {et~al.}(2010)\citenamefont {Guinea},
  \citenamefont {Katsnelson},\ and\ \citenamefont {Geim}}]{guinea2010energy}%
  \BibitemOpen
  \bibfield  {author} {\bibinfo {author} {\bibfnamefont {F.}~\bibnamefont
  {Guinea}}, \bibinfo {author} {\bibfnamefont {M.~I.}\ \bibnamefont
  {Katsnelson}},\ and\ \bibinfo {author} {\bibfnamefont {A.}~\bibnamefont
  {Geim}},\ }\href@noop {} {\bibfield  {journal} {\bibinfo  {journal} {Nature
  Physics}\ }\textbf {\bibinfo {volume} {6}},\ \bibinfo {pages} {30} (\bibinfo
  {year} {2010})}\BibitemShut {NoStop}%
\bibitem [{\citenamefont {Levy}\ \emph {et~al.}(2010)\citenamefont {Levy},
  \citenamefont {Burke}, \citenamefont {Meaker}, \citenamefont {Panlasigui},
  \citenamefont {Zettl}, \citenamefont {Guinea}, \citenamefont {Neto},\ and\
  \citenamefont {Crommie}}]{levy2010strain}%
  \BibitemOpen
  \bibfield  {author} {\bibinfo {author} {\bibfnamefont {N.}~\bibnamefont
  {Levy}}, \bibinfo {author} {\bibfnamefont {S.}~\bibnamefont {Burke}},
  \bibinfo {author} {\bibfnamefont {K.}~\bibnamefont {Meaker}}, \bibinfo
  {author} {\bibfnamefont {M.}~\bibnamefont {Panlasigui}}, \bibinfo {author}
  {\bibfnamefont {A.}~\bibnamefont {Zettl}}, \bibinfo {author} {\bibfnamefont
  {F.}~\bibnamefont {Guinea}}, \bibinfo {author} {\bibfnamefont {A.~C.}\
  \bibnamefont {Neto}},\ and\ \bibinfo {author} {\bibfnamefont {M.~F.}\
  \bibnamefont {Crommie}},\ }\href@noop {} {\bibfield  {journal} {\bibinfo
  {journal} {Science}\ }\textbf {\bibinfo {volume} {329}},\ \bibinfo {pages}
  {544} (\bibinfo {year} {2010})}\BibitemShut {NoStop}%
\bibitem [{\citenamefont {Jamadi}\ \emph {et~al.}(2020)\citenamefont {Jamadi},
  \citenamefont {Rozas}, \citenamefont {Salerno}, \citenamefont
  {Mili{\'c}evi{\'c}}, \citenamefont {Ozawa}, \citenamefont {Sagnes},
  \citenamefont {Lema{\^\i}tre}, \citenamefont {Le~Gratiet}, \citenamefont
  {Harouri}, \citenamefont {Carusotto} \emph {et~al.}}]{jamadi2020direct}%
  \BibitemOpen
  \bibfield  {author} {\bibinfo {author} {\bibfnamefont {O.}~\bibnamefont
  {Jamadi}}, \bibinfo {author} {\bibfnamefont {E.}~\bibnamefont {Rozas}},
  \bibinfo {author} {\bibfnamefont {G.}~\bibnamefont {Salerno}}, \bibinfo
  {author} {\bibfnamefont {M.}~\bibnamefont {Mili{\'c}evi{\'c}}}, \bibinfo
  {author} {\bibfnamefont {T.}~\bibnamefont {Ozawa}}, \bibinfo {author}
  {\bibfnamefont {I.}~\bibnamefont {Sagnes}}, \bibinfo {author} {\bibfnamefont
  {A.}~\bibnamefont {Lema{\^\i}tre}}, \bibinfo {author} {\bibfnamefont
  {L.}~\bibnamefont {Le~Gratiet}}, \bibinfo {author} {\bibfnamefont
  {A.}~\bibnamefont {Harouri}}, \bibinfo {author} {\bibfnamefont
  {I.}~\bibnamefont {Carusotto}}, \emph {et~al.},\ }\href@noop {} {\bibfield
  {journal} {\bibinfo  {journal} {Light: Science \& Applications}\ }\textbf
  {\bibinfo {volume} {9}},\ \bibinfo {pages} {144} (\bibinfo {year}
  {2020})}\BibitemShut {NoStop}%
\bibitem [{\citenamefont {Abbaszadeh}\ \emph {et~al.}(2017)\citenamefont
  {Abbaszadeh}, \citenamefont {Souslov}, \citenamefont {Paulose}, \citenamefont
  {Schomerus},\ and\ \citenamefont {Vitelli}}]{abbaszadeh2017sonic}%
  \BibitemOpen
  \bibfield  {author} {\bibinfo {author} {\bibfnamefont {H.}~\bibnamefont
  {Abbaszadeh}}, \bibinfo {author} {\bibfnamefont {A.}~\bibnamefont {Souslov}},
  \bibinfo {author} {\bibfnamefont {J.}~\bibnamefont {Paulose}}, \bibinfo
  {author} {\bibfnamefont {H.}~\bibnamefont {Schomerus}},\ and\ \bibinfo
  {author} {\bibfnamefont {V.}~\bibnamefont {Vitelli}},\ }\href@noop {}
  {\bibfield  {journal} {\bibinfo  {journal} {Physical review letters}\
  }\textbf {\bibinfo {volume} {119}},\ \bibinfo {pages} {195502} (\bibinfo
  {year} {2017})}\BibitemShut {NoStop}%
\bibitem [{\citenamefont {Brendel}\ \emph {et~al.}(2017)\citenamefont
  {Brendel}, \citenamefont {Peano}, \citenamefont {Painter},\ and\
  \citenamefont {Marquardt}}]{brendel2017pseudomagnetic}%
  \BibitemOpen
  \bibfield  {author} {\bibinfo {author} {\bibfnamefont {C.}~\bibnamefont
  {Brendel}}, \bibinfo {author} {\bibfnamefont {V.}~\bibnamefont {Peano}},
  \bibinfo {author} {\bibfnamefont {O.~J.}\ \bibnamefont {Painter}},\ and\
  \bibinfo {author} {\bibfnamefont {F.}~\bibnamefont {Marquardt}},\ }\href@noop
  {} {\bibfield  {journal} {\bibinfo  {journal} {Proceedings of the National
  Academy of Sciences}\ }\textbf {\bibinfo {volume} {114}},\ \bibinfo {pages}
  {E3390} (\bibinfo {year} {2017})}\BibitemShut {NoStop}%
\bibitem [{\citenamefont {Peri}\ \emph {et~al.}(2019)\citenamefont {Peri},
  \citenamefont {Serra-Garcia}, \citenamefont {Ilan},\ and\ \citenamefont
  {Huber}}]{peri2019axial}%
  \BibitemOpen
  \bibfield  {author} {\bibinfo {author} {\bibfnamefont {V.}~\bibnamefont
  {Peri}}, \bibinfo {author} {\bibfnamefont {M.}~\bibnamefont {Serra-Garcia}},
  \bibinfo {author} {\bibfnamefont {R.}~\bibnamefont {Ilan}},\ and\ \bibinfo
  {author} {\bibfnamefont {S.~D.}\ \bibnamefont {Huber}},\ }\href@noop {}
  {\bibfield  {journal} {\bibinfo  {journal} {Nature Physics}\ }\textbf
  {\bibinfo {volume} {15}},\ \bibinfo {pages} {357} (\bibinfo {year}
  {2019})}\BibitemShut {NoStop}%
\bibitem [{\citenamefont {Schomerus}\ and\ \citenamefont
  {Halpern}(2013)}]{schomerus2013parity}%
  \BibitemOpen
  \bibfield  {author} {\bibinfo {author} {\bibfnamefont {H.}~\bibnamefont
  {Schomerus}}\ and\ \bibinfo {author} {\bibfnamefont {N.~Y.}\ \bibnamefont
  {Halpern}},\ }\href@noop {} {\bibfield  {journal} {\bibinfo  {journal}
  {Physical review letters}\ }\textbf {\bibinfo {volume} {110}},\ \bibinfo
  {pages} {013903} (\bibinfo {year} {2013})}\BibitemShut {NoStop}%
\bibitem [{\citenamefont {Lled{\'o}}\ \emph {et~al.}(2022)\citenamefont
  {Lled{\'o}}, \citenamefont {Carusotto},\ and\ \citenamefont
  {Szymanska}}]{lledo2022polariton}%
  \BibitemOpen
  \bibfield  {author} {\bibinfo {author} {\bibfnamefont {C.}~\bibnamefont
  {Lled{\'o}}}, \bibinfo {author} {\bibfnamefont {I.}~\bibnamefont
  {Carusotto}},\ and\ \bibinfo {author} {\bibfnamefont {M.}~\bibnamefont
  {Szymanska}},\ }\href@noop {} {\bibfield  {journal} {\bibinfo  {journal}
  {SciPost Physics}\ }\textbf {\bibinfo {volume} {12}},\ \bibinfo {pages} {068}
  (\bibinfo {year} {2022})}\BibitemShut {NoStop}%
\bibitem [{\citenamefont {Li}\ \emph {et~al.}(2008)\citenamefont {Li},
  \citenamefont {White}, \citenamefont {O’Faolain}, \citenamefont
  {Gomez-Iglesias},\ and\ \citenamefont {Krauss}}]{li2008systematic}%
  \BibitemOpen
  \bibfield  {author} {\bibinfo {author} {\bibfnamefont {J.}~\bibnamefont
  {Li}}, \bibinfo {author} {\bibfnamefont {T.~P.}\ \bibnamefont {White}},
  \bibinfo {author} {\bibfnamefont {L.}~\bibnamefont {O’Faolain}}, \bibinfo
  {author} {\bibfnamefont {A.}~\bibnamefont {Gomez-Iglesias}},\ and\ \bibinfo
  {author} {\bibfnamefont {T.~F.}\ \bibnamefont {Krauss}},\ }\href@noop {}
  {\bibfield  {journal} {\bibinfo  {journal} {Optics express}\ }\textbf
  {\bibinfo {volume} {16}},\ \bibinfo {pages} {6227} (\bibinfo {year}
  {2008})}\BibitemShut {NoStop}%
\bibitem [{\citenamefont {Juki{\'c}}\ \emph {et~al.}(2012)\citenamefont
  {Juki{\'c}}, \citenamefont {Buljan}, \citenamefont {Lee}, \citenamefont
  {Joannopoulos},\ and\ \citenamefont {Solja{\v{c}}i{\'c}}}]{jukic2012flat}%
  \BibitemOpen
  \bibfield  {author} {\bibinfo {author} {\bibfnamefont {D.}~\bibnamefont
  {Juki{\'c}}}, \bibinfo {author} {\bibfnamefont {H.}~\bibnamefont {Buljan}},
  \bibinfo {author} {\bibfnamefont {D.-H.}\ \bibnamefont {Lee}}, \bibinfo
  {author} {\bibfnamefont {J.~D.}\ \bibnamefont {Joannopoulos}},\ and\ \bibinfo
  {author} {\bibfnamefont {M.}~\bibnamefont {Solja{\v{c}}i{\'c}}},\ }\href@noop
  {} {\bibfield  {journal} {\bibinfo  {journal} {Optics Letters}\ }\textbf
  {\bibinfo {volume} {37}},\ \bibinfo {pages} {5262} (\bibinfo {year}
  {2012})}\BibitemShut {NoStop}%
\bibitem [{\citenamefont {Yang}\ \emph {et~al.}(2023)\citenamefont {Yang},
  \citenamefont {Roques-Carmes}, \citenamefont {Kooi}, \citenamefont {Tang},
  \citenamefont {Beroz}, \citenamefont {Mazur}, \citenamefont {Kaminer},
  \citenamefont {Joannopoulos},\ and\ \citenamefont
  {Solja{\v{c}}i{\'c}}}]{yang2023photonic}%
  \BibitemOpen
  \bibfield  {author} {\bibinfo {author} {\bibfnamefont {Y.}~\bibnamefont
  {Yang}}, \bibinfo {author} {\bibfnamefont {C.}~\bibnamefont {Roques-Carmes}},
  \bibinfo {author} {\bibfnamefont {S.~E.}\ \bibnamefont {Kooi}}, \bibinfo
  {author} {\bibfnamefont {H.}~\bibnamefont {Tang}}, \bibinfo {author}
  {\bibfnamefont {J.}~\bibnamefont {Beroz}}, \bibinfo {author} {\bibfnamefont
  {E.}~\bibnamefont {Mazur}}, \bibinfo {author} {\bibfnamefont
  {I.}~\bibnamefont {Kaminer}}, \bibinfo {author} {\bibfnamefont {J.~D.}\
  \bibnamefont {Joannopoulos}},\ and\ \bibinfo {author} {\bibfnamefont
  {M.}~\bibnamefont {Solja{\v{c}}i{\'c}}},\ }\href@noop {} {\bibfield
  {journal} {\bibinfo  {journal} {Nature}\ }\textbf {\bibinfo {volume} {613}},\
  \bibinfo {pages} {42} (\bibinfo {year} {2023})}\BibitemShut {NoStop}%
\bibitem [{\citenamefont {Guglielmon}\ \emph {et~al.}(2021)\citenamefont
  {Guglielmon}, \citenamefont {Rechtsman},\ and\ \citenamefont
  {Weinstein}}]{guglielmon2021Landau}%
  \BibitemOpen
  \bibfield  {author} {\bibinfo {author} {\bibfnamefont {J.}~\bibnamefont
  {Guglielmon}}, \bibinfo {author} {\bibfnamefont {M.~C.}\ \bibnamefont
  {Rechtsman}},\ and\ \bibinfo {author} {\bibfnamefont {M.~I.}\ \bibnamefont
  {Weinstein}},\ }\href@noop {} {\bibfield  {journal} {\bibinfo  {journal}
  {Physical Review A}\ }\textbf {\bibinfo {volume} {103}},\ \bibinfo {pages}
  {013505} (\bibinfo {year} {2021})}\BibitemShut {NoStop}%
\bibitem [{\citenamefont {Men}\ \emph {et~al.}(2014)\citenamefont {Men},
  \citenamefont {Lee}, \citenamefont {Freund}, \citenamefont {Peraire},\ and\
  \citenamefont {Johnson}}]{men2014robust}%
  \BibitemOpen
  \bibfield  {author} {\bibinfo {author} {\bibfnamefont {H.}~\bibnamefont
  {Men}}, \bibinfo {author} {\bibfnamefont {K.~Y.}\ \bibnamefont {Lee}},
  \bibinfo {author} {\bibfnamefont {R.~M.}\ \bibnamefont {Freund}}, \bibinfo
  {author} {\bibfnamefont {J.}~\bibnamefont {Peraire}},\ and\ \bibinfo {author}
  {\bibfnamefont {S.~G.}\ \bibnamefont {Johnson}},\ }\href@noop {} {\bibfield
  {journal} {\bibinfo  {journal} {Optics express}\ }\textbf {\bibinfo {volume}
  {22}},\ \bibinfo {pages} {22632} (\bibinfo {year} {2014})}\BibitemShut
  {NoStop}%
\bibitem [{\citenamefont {Vuckovic}\ \emph {et~al.}(2002)\citenamefont
  {Vuckovic}, \citenamefont {Loncar}, \citenamefont {Mabuchi},\ and\
  \citenamefont {Scherer}}]{vuckovic2002optimization}%
  \BibitemOpen
  \bibfield  {author} {\bibinfo {author} {\bibfnamefont {J.}~\bibnamefont
  {Vuckovic}}, \bibinfo {author} {\bibfnamefont {M.}~\bibnamefont {Loncar}},
  \bibinfo {author} {\bibfnamefont {H.}~\bibnamefont {Mabuchi}},\ and\ \bibinfo
  {author} {\bibfnamefont {A.}~\bibnamefont {Scherer}},\ }\href@noop {}
  {\bibfield  {journal} {\bibinfo  {journal} {IEEE Journal of Quantum
  Electronics}\ }\textbf {\bibinfo {volume} {38}},\ \bibinfo {pages} {850}
  (\bibinfo {year} {2002})}\BibitemShut {NoStop}%
\bibitem [{\citenamefont {Bendsoe}\ and\ \citenamefont
  {Sigmund}(2003)}]{bendsoe2003topology}%
  \BibitemOpen
  \bibfield  {author} {\bibinfo {author} {\bibfnamefont {M.~P.}\ \bibnamefont
  {Bendsoe}}\ and\ \bibinfo {author} {\bibfnamefont {O.}~\bibnamefont
  {Sigmund}},\ }\href@noop {} {\emph {\bibinfo {title} {Topology optimization:
  theory, methods, and applications}}}\ (\bibinfo  {publisher} {Springer
  Science \& Business Media},\ \bibinfo {year} {2003})\BibitemShut {NoStop}%
\bibitem [{\citenamefont {Barik}\ \emph {et~al.}(2016)\citenamefont {Barik},
  \citenamefont {Miyake}, \citenamefont {DeGottardi}, \citenamefont {Waks},\
  and\ \citenamefont {Hafezi}}]{barik2016two}%
  \BibitemOpen
  \bibfield  {author} {\bibinfo {author} {\bibfnamefont {S.}~\bibnamefont
  {Barik}}, \bibinfo {author} {\bibfnamefont {H.}~\bibnamefont {Miyake}},
  \bibinfo {author} {\bibfnamefont {W.}~\bibnamefont {DeGottardi}}, \bibinfo
  {author} {\bibfnamefont {E.}~\bibnamefont {Waks}},\ and\ \bibinfo {author}
  {\bibfnamefont {M.}~\bibnamefont {Hafezi}},\ }\href@noop {} {\bibfield
  {journal} {\bibinfo  {journal} {New Journal of Physics}\ }\textbf {\bibinfo
  {volume} {18}},\ \bibinfo {pages} {113013} (\bibinfo {year}
  {2016})}\BibitemShut {NoStop}%
\bibitem [{\citenamefont {Neto}\ \emph {et~al.}(2009)\citenamefont {Neto},
  \citenamefont {Guinea}, \citenamefont {Peres}, \citenamefont {Novoselov},\
  and\ \citenamefont {Geim}}]{neto2009electronic}%
  \BibitemOpen
  \bibfield  {author} {\bibinfo {author} {\bibfnamefont {A.~C.}\ \bibnamefont
  {Neto}}, \bibinfo {author} {\bibfnamefont {F.}~\bibnamefont {Guinea}},
  \bibinfo {author} {\bibfnamefont {N.~M.}\ \bibnamefont {Peres}}, \bibinfo
  {author} {\bibfnamefont {K.~S.}\ \bibnamefont {Novoselov}},\ and\ \bibinfo
  {author} {\bibfnamefont {A.~K.}\ \bibnamefont {Geim}},\ }\href@noop {}
  {\bibfield  {journal} {\bibinfo  {journal} {Reviews of modern physics}\
  }\textbf {\bibinfo {volume} {81}},\ \bibinfo {pages} {109} (\bibinfo {year}
  {2009})}\BibitemShut {NoStop}%
\bibitem [{\citenamefont {Fefferman}\ and\ \citenamefont
  {Weinstein}(2012)}]{fefferman2012honeycomb}%
  \BibitemOpen
  \bibfield  {author} {\bibinfo {author} {\bibfnamefont {C.~L.}\ \bibnamefont
  {Fefferman}}\ and\ \bibinfo {author} {\bibfnamefont {M.~I.}\ \bibnamefont
  {Weinstein}},\ }\href@noop {} {\bibfield  {journal} {\bibinfo  {journal}
  {Journal of the American Mathematical Society}\ }\textbf {\bibinfo {volume}
  {25}},\ \bibinfo {pages} {1169} (\bibinfo {year} {2012})}\BibitemShut
  {NoStop}%
\bibitem [{\citenamefont {Lee-Thorp}\ \emph {et~al.}(2019)\citenamefont
  {Lee-Thorp}, \citenamefont {Weinstein},\ and\ \citenamefont
  {Zhu}}]{Dirac2DPhotonic}%
  \BibitemOpen
  \bibfield  {author} {\bibinfo {author} {\bibfnamefont {J.~P.}\ \bibnamefont
  {Lee-Thorp}}, \bibinfo {author} {\bibfnamefont {M.~I.}\ \bibnamefont
  {Weinstein}},\ and\ \bibinfo {author} {\bibfnamefont {Y.}~\bibnamefont
  {Zhu}},\ }\href@noop {} {\bibfield  {journal} {\bibinfo  {journal} {Archive
  for Rational Mechanics and Analysis}\ }\textbf {\bibinfo {volume} {232}},\
  \bibinfo {pages} {1} (\bibinfo {year} {2019})}\BibitemShut {NoStop}%
\bibitem [{\citenamefont {Minkov}\ \emph {et~al.}(2020)\citenamefont {Minkov},
  \citenamefont {Williamson}, \citenamefont {Andreani}, \citenamefont {Gerace},
  \citenamefont {Lou}, \citenamefont {Song}, \citenamefont {Hughes},\ and\
  \citenamefont {Fan}}]{Legume}%
  \BibitemOpen
  \bibfield  {author} {\bibinfo {author} {\bibfnamefont {M.}~\bibnamefont
  {Minkov}}, \bibinfo {author} {\bibfnamefont {I.~A.}\ \bibnamefont
  {Williamson}}, \bibinfo {author} {\bibfnamefont {L.~C.}\ \bibnamefont
  {Andreani}}, \bibinfo {author} {\bibfnamefont {D.}~\bibnamefont {Gerace}},
  \bibinfo {author} {\bibfnamefont {B.}~\bibnamefont {Lou}}, \bibinfo {author}
  {\bibfnamefont {A.~Y.}\ \bibnamefont {Song}}, \bibinfo {author}
  {\bibfnamefont {T.~W.}\ \bibnamefont {Hughes}},\ and\ \bibinfo {author}
  {\bibfnamefont {S.}~\bibnamefont {Fan}},\ }\href@noop {} {\bibfield
  {journal} {\bibinfo  {journal} {Acs Photonics}\ }\textbf {\bibinfo {volume}
  {7}},\ \bibinfo {pages} {1729} (\bibinfo {year} {2020})}\BibitemShut
  {NoStop}%
\bibitem [{\citenamefont {Bender}\ and\ \citenamefont
  {Orszag}(1999)}]{bender1999advanced}%
  \BibitemOpen
  \bibfield  {author} {\bibinfo {author} {\bibfnamefont {C.~M.}\ \bibnamefont
  {Bender}}\ and\ \bibinfo {author} {\bibfnamefont {S.~A.}\ \bibnamefont
  {Orszag}},\ }\href@noop {} {\emph {\bibinfo {title} {Advanced mathematical
  methods for scientists and engineers I: Asymptotic methods and perturbation
  theory}}},\ Vol.~\bibinfo {volume} {1}\ (\bibinfo  {publisher} {Springer
  Science \& Business Media},\ \bibinfo {year} {1999})\BibitemShut {NoStop}%
\end{thebibliography}%

\end{document}